\documentclass[10pt]{article}
\usepackage{amsmath, amsthm, amssymb}
\usepackage{color}
\usepackage{listings}
\usepackage{a4wide}
\usepackage{courier}
\usepackage{comment}
\usepackage{amsfonts}
\usepackage{graphicx}
\usepackage[utf8]{inputenc}
\usepackage[hang,small]{caption}
\usepackage[numbers,sort]{natbib}

\begin{titlepage}
\title{Adaptive Multi-GPU Exchange Monte Carlo for the 3D Random Field Ising Model}
\author{C. A. Navarro$^{1,2}$, Wei Huang$^{3}$, Youjin Deng $^{3,4}$         \\\\
         $^1$ \small Department of Computer Science, Universidad de Chile, Santiago, Chile.\\
         $^2$ \small Centro de Estudios Científicos (CECs), Valdivia, Chile.\\
		 $^3$ \small Hefei National Laboratory for Physical Sciences at Microscale, Department of Modern Physics,\\
	 		  \small University of Science and Technology of China, Hefei, 230027, China.\\
         $^4$ \small State Key Laboratory of Theoretical Physics, Institute of Theoretical Physics,\\
              \small Chinese Academy of Sciences, Beijing 100190, China.
}
\date{\today}
\end{titlepage}

\newcommand{\bibtexdb}{\string~/Dropbox/bibtexdb/bibliography.bib}
\begin{document}
\maketitle

\begin{abstract}
We present an adaptive multi-GPU Exchange Monte Carlo implementation designed for
the simulation of the 3D Random Field Model. The original algorithm is re-designed
based on a two-level parallelization scheme that allows the method to scale its
performance in the presence of faster and GPUs as well as multiple GPUs. The set of
temperatures is adapted according to the exchange rate observed from short trial
runs, leading to an increased exchange rate at zones where the exchange process is
sporadic.  Performance results show that parallel tempering is an ideal strategy for
being implemented on the GPU, and runs between one to two orders of magnitude with
respect to a single-core CPU version, with multi-GPU scaling being approximately
$99\%$ efficient. The results obtained extend the possibilities of simulation to
sizes of $L = 32, 64$ for a workstation with two GPUs.
\end{abstract}

\section{Introduction}
Monte Carlo methods have become a convenient strategy for simulating finite size spin
lattices towards equilibrium and to perform average measurements of physical
observables.  Classic spin models such as Ising \cite{ising_1925, ising_intro} and
Potts \cite{Potts:479866} are usually simulated with the Metropolis-Hastings
algorithm \cite{metropolis53, hastings70}, cluster \cite{PhysRevLett.58.86,
PhysRevLett.62.361, PhysRevB.43.10617} or worm algorithm \cite{Prokof'ev}, with the
last two being more efficient reaching equilibrium near the critical temperature
$T_c$ \cite{sokal_csd_now}.

\begin{figure}[ht!]
    \includegraphics[scale=1.8]{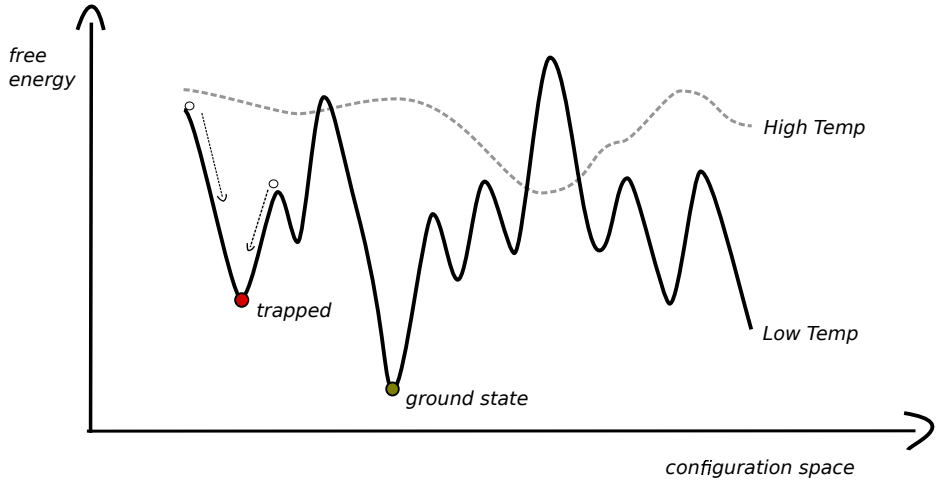}
    \centering
    \caption{At low temperature, classic algorithms fail to reach the ground state
    for systems with quenched disorder.}
    \label{fig_roughlandscape_paper}
\end{figure}

For systems with \textit{quenched disorder} such as the Spin Glass and the Random Field Ising Model
(RFIM) we can find that classic algorithms are not efficient anymore in the low
temperature regime. In this work we are interested in studying the 3D RFIM, which
introduces a disordered magnetic field $\{h_1, h_2, \cdots, h_n\}$ of strength $h$, to the Hamiltonian:
\begin{equation} 
\mathcal{H} = -\sum_{\langle i,j \rangle}{J s_is_j} -
h\sum_{i}{h_i s_i} 
\end{equation} 
The reason why classic MCMC algorithms fail is because systems with quenched disorder
present an adverse energy landscape, making the simulation with a classic MCMC algorithm to
easily become trapped in a local minimum, thus never reaching the ground state of the
system. Figure
\ref{fig_roughlandscape_paper} illustrates the problem.
The problem occurs at low temperatures, \textit{i.e.}, $T \le T_c$, which is a
required zone to explore if one is studying phase transitions.  Instead of simulating
the system with one instance of the lattice, as in classic algorithms, one can
simulate $R$ replicas at different temperatures and exchange information among them,
to eventually overcome the local minimum problem \cite{JPSJ.65.1604,
PhysRevLett.57.2607}. By exchanging information between replicas from time to time,
information from the high temperature regime arrives to the replicas at low
temperature, \textit{shaking} the system and providing the opportunity to escape the
local minimum. The algorithm based on this principle is known as the \textit{Exchange
Monte Carlo} method, also as \textit{Parallel Tempering}, and it is one of the most
used algorithms for simulating systems with quenched disorder.  The notion of
exchanging replicas for studying systems with quenched disorder was first introduced
by Swendsen and Wang in 1986 \cite{PhysRevLett.57.2607} and then extended by Geyer in
1991 \cite{Geyer:1991}.  Hukushima and Nemoto presented in 1996 the full method as it
is known today \cite{JPSJ.65.1604}.

The replica based approach has been used to simulate 2D and 3D spin glasses
\cite{JPSJ.65.1604, 0305-4608-5-5-017, ogielski1985critical,
katzgraber2006universality, rieger1996critical}.  While it is true that replica based
methods overcome the main difficulty of the Monte Carlo simulation in disordered
systems, the computational cost is still considered a problem, since it requires at
least $\Omega(R\cdot L^d)$ per time step, with $L$ being the linear size of the
lattice and $d$ the number of dimensions. Furthermore, the number of simulation steps
required to reach the ground state is in the order of millions, and increases as the
lattice gets larger.  In this work we are interested in the particular case of
simulating the 3D Ising Random Field model at sizes $L = \{8, 16, 32, 64\}$ using the
Exchange Monte Carlo method adapted to massively parallel architectures such as the
GPU. 

The fast expansion of \textit{massively parallel architectures} \cite{navhitmat2014}
provides an opportunity to further improve the running time of data-parallel
algorithms \cite{magoules:proceedings-auth:48, magoules:proceedings-auth:50,
magoules:journal-auth:53};  \textit{Parallel Tempering} (PT) in this case. The GPU
architecture, being a \textit{massively parallel architecture}, can easily perform an
order of magnitude faster than a CPU. Moreover, the GPU is more energy efficient and
cheaper than classic clusters and supercomputers based on CPU hardware. But in order
to achieve such level of performance, GPU-based algorithms need to be carefully
designed and implemented, presenting a great challenge on the computational side.
This computational challenge is what motivates our work.

In this work, we propose a multi-GPU method for modern GPU architectures
for the simulation of the \textit{3D Random Field Ising Model}.  The
implementation uses two levels of parallelism; (1) \textit{spin parallelism}
that scales in the presence of faster GPUs, and (2) \textit{replica parallelism}
that scales in the presence of multiple GPUs. Both levels, when combined
together, provide a substantial boost in performance that allows the study of
problems that were too large in the past for a conventional CPU implementation,
such as $L = \{8, 16, 32, 64\}$.  In addition to the parallelization strategy, we also
propose a new temperature selection scheme, based on recursive insertion of
points, to improve the exchange rate at the zones where exchange is often less
frequent. Physical results have been included for the 3D random field model at
sizes $L = \{8, 16, 32, 64\}$.

The rest of the Chapter is organized as follows: Section \ref{sec_multigpu_relatedwork}
presents the related work regarding parallel implementations of the Exchange
Monte Carlo method. In Section \ref{sec_GPUapproach} we point out the levels of
parallelism present in the Exchange Monte Carlo method, and explain the
multi-GPU method in its two levels. Section \ref{sec_ATS} presents the
\textit{Adaptive Temperature} strategy we use for choosing the temperature
distribution and show how it can improve the results of the simulation as well
as reduce simulation time.  In Section \ref{sec_multigpu_performance} we show the
experimental performance results, which consist of comparison against a CPU
implementation, performance scaling under different GPUs as well as under one
and two GPUs, and results on the improvement provided by the adaptation
technique, compared to a simulation without the approach. Section
\ref{sec_multigpu_adaptive_exchange} we present the exchange rate of the adaptive 
strategy and compare it to other homogeneous approaches. In Section
\ref{sec_physical} we present some physical results on the 3D Ising Random
Field, for sizes $L = \{8, 16, 32, 64\}$.  Finally, in Section
\ref{sec_multigpu_conclusions} we discuss our results and conclude our work.

\section{Related Work}
\label{sec_multigpu_relatedwork}
Several works have shown the benefits of GPU-based implementations of MCMC algorithms
for spin systems. The Metropolis-Hastings algorithm has been efficiently re-designed
as a GPU algorithm for both 2D and 3D lattices \cite{ferrero2012, Preis20094468, Block20101549, Lulli2015}. 
The parallelization strategy is usually based on the
\textit{checkerboard} decomposition of the problem domain, where black and white
spins are simulated in a two-step parallel computation. Although the checkerboard
method violates \textit{detailed balance}, it still obeys the \textit{global balance}
condition which is sufficient to ensure convergence of the stochastic process.  M.
Weigel proposed the \textit{double checkerboard} strategy (see Figure
\ref{fig_multigpu_doublecheckerboard2D}), that takes advantage of the GPU's shared
memory \cite{weigel2011simulating, weigel2012performance, weigel2012simulating} for
doing partial Metropolis sweeps entirely in cache. In the work of Lulli \textit{et.
al.}, the authors propose to reorganize the lattice in alternating slices in order to 
achieve efficient memory access patterns. 

For cluster algorithms, recent
works have proposed single and multiple GPU implementations, for both Ising and Potts
models \cite{Komura201340, Komura:2012:GSA:2088552.2088820,
DBLP:journals/cphysics/KomuraO12, weigel:10b}.  For the case of the Swendsen and Wang
algorithm, which is a multi-cluster one, some use a parallel labeling strategy based
on the work of Hawick \textit{et. al.} \cite{Hawick:2010:PGC:1869135.1869229}. The
cluster work of Weigel uses an approach based on self-labeling with hierarchical
sewing and label relaxation \cite{weigel:10b}.  A study on the parallelism of the
Worm algorithm has been reported by Delgado et al. \cite{ydelgado11}. The authors
conclude that an efficient GPU parallelization is indeed hard because very few worms
stay alive at a given time. 

Two GPU-based implementations of the Exchange Monte Carlo method have been
proposed. The first one was proposed by Weigel for 2D Spin Glasses
\cite{weigel2012performance}, in which the author treats all the replicas of the
system as one large lattice, therefore additional replicas in practice turn out
to be additional thread blocks. The second work is by Ye Fang \textit{et. al}
\cite{Fang20142467} and they propose a fast multi-GPU implementation for studying the
3D Spin Glass. In their work, the authors propose to keep the replicas in shared
memory instead of global memory. This modification provides a performance memory
accesses that are an order of magnitude faster than global memory ones, but limits
the lattice size to the size of the shared memory, which for todays GPUs it means 3D
lattices of size $L \le 16$.
Katzgraber \textit{et. al.} proposed a method for improving the temperature
set in the Exchange Monte Carlo method \cite{1742-5468-2006-03-P03018}. The 
strategy is based on keeping a histogram record of the number of round trips 
of each replica (\textit{i.e.}, the number of times a replica travels from $T_{min}$
to $T_{max}$ and vice versa), and improving the locations where this value is the lowest.
Another strategy was presented by Bittner \textit{et. al.}, where they 
propose a method for obtaining a good set of temperatures and also they 
propose to set the number of lattice sweeps according to the auto-correlation 
time observed \cite{PhysRevLett.101.130603}. 

To the best of our knowledge, there is still room for additional improvements
regarding GPU implementations for the Exchange Monte Carlo method, such as using
concurrent kernel launches, extending the double checkerboard strategy to 3D,
optimal 3D thread blocks, global-memory multi-GPU partitions, and low level
optimizations for the case of the 3D Random Field model, among others. In 
relation to choosing the temperature set, it is still possible to explore 
different adaptive strategies based on recursive algorithms. In the 
next section we present our parallel implementation of the Exchange Monte Carlo 
method as well as our strategy for choosing an efficient temperature set.

\section{Multi-GPU approach} 
\label{sec_GPUapproach} 
For a multi-GPU approach, we analyze the \textit{Exchange Monte Carlo} to 
find out how many levels of parallelism exist.

\subsection{Parallelism in the Exchange Monte Carlo method}
The \textit{Exchange Monte Carlo} method is an
algorithm for simulating systems with \textit{quenched disorder}. The notion of
exchanging replicas was first introduced by Swendsen and Wang in 1986
\cite{PhysRevLett.57.2607}, later extended by Geyer in 1991 \cite{Geyer:1991}. In
1996, Hukushima and Nemoto formulated the algorithm as is it known today
\cite{JPSJ.65.1604}.  The algorithm has become widely known for its efficiency at
simulating Spin Glasses, and for its simplicity in its definition. Due to its general
definition, the algorithm can be applied with no difficulties to other models
different from the spin glass model such as the Random Field Ising Model (RFIM), 
which is the model of study in this work.

\begin{figure}[ht!]
    \includegraphics[scale=0.57]{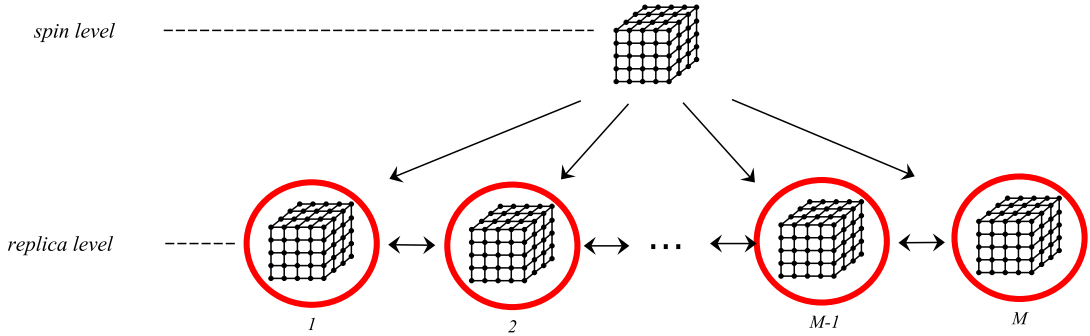}
    \centering
    \caption{The two levels of parallelization available in the exchange Monte
    Carlo method.}
    \label{fig_twolevels}
\end{figure}

The algorithm works with $M$ \textit{replicas} $\mathcal{X} = \{X_1, X_2, ..., X_M\}$ of the
system with each one at a different temperature.  The main steps for one disorder realization of
our parallel \textit{Exchange Monte Carlo} algorithm, for the case of the Ising
Random Field model, are the following: 
\begin{enumerate} 
\item{Choose $M$ different temperatures $\{\beta_1, \beta_2, ..., \beta_M\}$ where $\beta = 1/T$.} 
\item{Choose an arbitrary random magnetic field $H = \{h_1, h_2, h_3..., h_{|V|}\}$ with $h_i = rand(\pm1)$. 
    This instance $H$ of disorder is used for
the entire simulation by all $M$ replicas.} 
\item{Set an arbitrary spin configuration to each one of
the $M$ replicas and assign the corresponding temperature, \textit{i.e.}, $X_i
\leftarrow \beta_i$.}
\item{\textbf{[Parallel]} Simulate each replica simultaneously and independently in the Random Field
Model for $p$ parallel tempering moves, using $H$ for all replicas and a highly
parallel MCMC algorithm such as Metropolis-Hastings. At each
parallel tempering move, exchange the \textit{odd \textbf{xor} even} configurations
$X_i$ with their next neighbor $X_{i+1}$, with probability
    \begin{equation}
        W(X_i, \beta_i | X_{i+1}, \beta_{i+1}) = \begin{cases}
        1 & \text{for $\Delta < 0$}\\
        \mathit{e}^{\Delta} & \text{for $\Delta > 0$}
        \end{cases}
	\end{equation}
	where $\Delta = (\beta_{i+1} - \beta_{i})(\mathcal{H}(X_{i+1}) - \mathcal{H}(X_i))$.
    Choosing odd or even depends if the $j$-th exchange is odd or even, respectively. 
}
\end{enumerate}
The algorithm itself is inherently \textit{data-parallel} for step (4)
and provides a sufficient number of data elements for a GPU
implementation. In fact, there are two levels of parallelism that can be
exploited; (1) \textit{spin parallelism} and (2) \textit{replica level
parallelism}.  In \textit{spin parallelism} the challenge is to come up
with a classic MCMC method that can take full advantage of the GPU parallel
power.  For this, we use a GPU-based Metropolis-Hastings implementation
optimized for 3D lattices. For (2), the problem seems \textit{pleasingly
parallel} for a multi-GPU implementation, however special care must be put at
the exchange phase, since there is a potential memory bottleneck due to the
distributed memory for a multi-GPU approach. Figure \ref{fig_twolevels}
illustrates the two levels of parallelism and their organization.

\subsection{Spin Level Parallelism}
\textit{Spin parallelism} corresponds to the parallel simulation of the spins of a single
replica and we propose to handle it by using a single CUDA kernel based on
the \textit{double checkerboard} idea, proposed by M. Weigel
\cite{weigel2012simulating, weigel2012performance}. A \textit{double
checkerboard} approach allows the efficient simulation of spin systems using
coalesced memory accesses, as well as the option to choose multiple partial
sweeps in the same kernel at a much higher performance because of the GPU's
shared memory. In the original works by Weigel, the optimizations are only
described and implemented for the 2D Ising Spin Glass
\cite{weigel2012simulating, weigel2012performance}, but the author mentions that
the ideas can be extended to 3D by using a more elaborate thread indexing
scheme. The \textit{spin parallelization} implementation of this work is a
\textit{double 3D checkerboard} and corresponds to the extension mentioned by
the author. 

\begin{figure}[ht!]
    \includegraphics[scale=0.14]{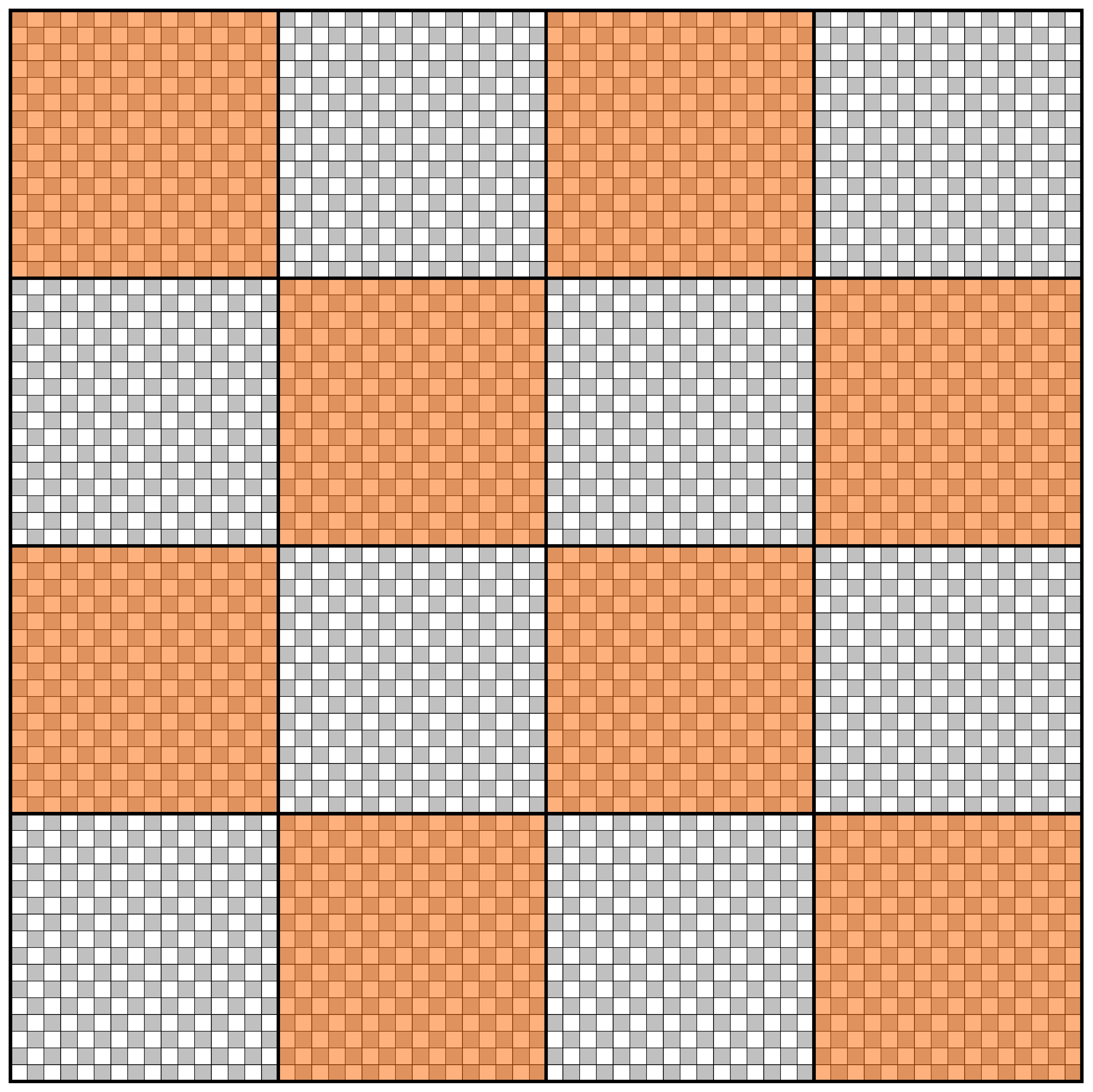}
    \centering
    \caption{A double 2D checkerboard of $D \times D = 4 \times 4$
    tiles, each one containing $T \times T = 16 \times 16$ spins.}
    \label{fig_multigpu_doublecheckerboard2D}
\end{figure}

A \textit{double checkerboard} is a \textit{two-fold} Metropolis-Hastings
simulation strategy that is composed of several fine grained checkerboards
organized into one coarse checkerboard.  The case of 2D is illustrated in Figure
\ref{fig_multigpu_doublecheckerboard2D}.

The double checkerboard method starts by simulating all spins from the orange
tiles of $T \times T$ spins, doing a local two-step \textit{black-white}
simulation on the fine grained checkerboard of $T \times T$. The tile
sub-checkerboard is loaded entirely in GPU shared memory, including a halo of
spins that reside on the neighbor tiles of the opposite color. After all orange
tiles are finished, the same is done for the white tiles. Two kernel executions
are needed to fully synchronize all threads when changing tile color. While it is
true that only $L^3 / 4$ spins are being simulated at a given time, its
advantage is that the memory access pattern is fully coalesced and the spin flip
is performed in shared memory.  

\begin{figure}[ht!]
    \includegraphics[scale=0.6]{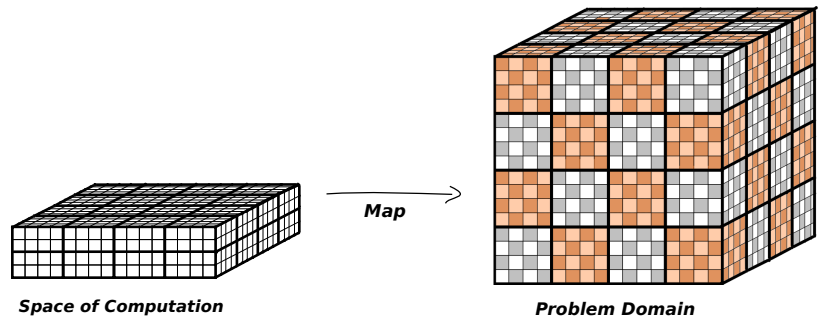}
    \centering
    \caption{A space of computation of size $L^3/4$ threads is necessary and sufficient for simulating
    the spins in parallel using a double 3D checkerboard approach.}
    \label{fig_map_doublecheckerboard3D}
\end{figure}

In order to create a \textit{double 3D checkerboard}, we convert both the fine
and coarse grained checkerboards to 3D. For this, we use the fact that a 3D
checkerboard can be generated by using alternated 2D checkerboard layers stacked
over a third dimension. For a given point $p = (x,y,z)$ in 3D discrete space, its
alternation value $A = \{0,1\}$ is defined as
\begin{equation}
A = (p_x + p_y + p_z) \mod 2
\label{eq_multigpu_alternator}
\end{equation}
Indeed one could build a \textit{space of computation} of the size of the whole
lattice, launch the kernel and compute the value of $A$ using expression
(\ref{eq_multigpu_alternator}), but this approach would be inefficient because
$3L^3/4$ threads would remain idle waiting for its turn in the checkerboard
process. Instead, we use a space of computation composed of $L^3 / 4$ threads; 
$T^3 / 2$ threads per block and $D^3 / 2$ blocks, where $D$ is the number of
tiles per dimension. Figure \ref{fig_map_doublecheckerboard3D} illustrates 
a space of computation of $L^3 / 4$ threads being sufficient for handling a 
double 3D checkerboard.

For the halos it is important to consider what percentage of the spins loaded
into the GPU shared memory will actually be halo spins. Halo spins are more expensive
to load into shared memory than internal tile spins due to the unaligned memory
access pattern, therefore one would want the minimum number of halo spins in the
shared memory.  Considering that actual GPU architectures have a constant warp of
threads $w$ and require a constant block volume $V$ to be specified, finding the minimum halo
is equivalent to solve the optimization problem of minimizing the surface of a closed box
with dimensions $w \times x \times y$ and constant volume $V$. The objective function
to minimize is the surface expression
\begin{equation}
S = 2(wy + xy + wx)
\end{equation}
subject to $V = wxy$, from where we can rewrite $S$ in one variable.
Setting the derivative of $S$ to zero leads to
\begin{equation}
\frac{\partial S}{\partial x} = 2(w - V/x^2) = 0
\end{equation}
where we finally get the solution: $x = y = \sqrt{V / w}$. Considering that the number of
spins inside a tile is double the number of threads (because of the checkerboard
approach) and that the maximum number of threads in a block is $B_{max} = 1024$ for 
actual GPUs, we have that $V = 2|B_{max}|$. With this, $x = y = 8$ 
and the optimal block to use is $B_{opt} = (32, 4, 8)$.

\subsection{Replica Level Parallelism}
Replica level parallelism is based on the combination of concurrent kernel
execution from modern GPUs and coarse parallelism from the multi-GPU computing.
In modern GPU architectures, one can launch multiple kernels in the same GPU and
let the driver scheduler handle the physical resources to execute these kernels
concurrently for that GPU. Starting from the Kepler GPU architecture, it is
possible to launch up to $32$ kernels concurrently on a single GPU.  The idea is
to divide the $M$ replicas into $k$ available GPUs and simulate $m = \lceil M/k
\rceil$ replicas concurrently for each GPU. It is possible that $m > 32$, but it
is not a problem because the GPU can handle the exceeding kernels automatically
with very small overhead, by using an internal execution queue.  With the new
approach, the new number of replicas becomes 
\begin{equation} 
M' = D\cdot m \le M + D
\end{equation} 
where $D$ is the number of GPUs used in the multi-GPU computation. By using $M'$ 
replicas instead of $M$, we guarantee a balanced parallel computation for all GPUs 
and at the same time the extra replicas help to produce a better result. One
assumes that $D \le M$. 

In a multi-GPU approach, the shared memory scenario transforms into a
distributed scenario that must be handled using global indexing of the local
memory regions.  For each GPU, there is a region of memory allocated for the $m$
replicas. For any GPU $D_i$, the global index for its \textit{left-most} replica
is $D_{i}^L = m \cdot i$ while the global index for its \textit{right-most}
replica is $D_{i}^R = m \cdot i + m-1 = m(i+1) - 1$. 

\begin{figure}[ht!]
    \includegraphics[scale=0.18]{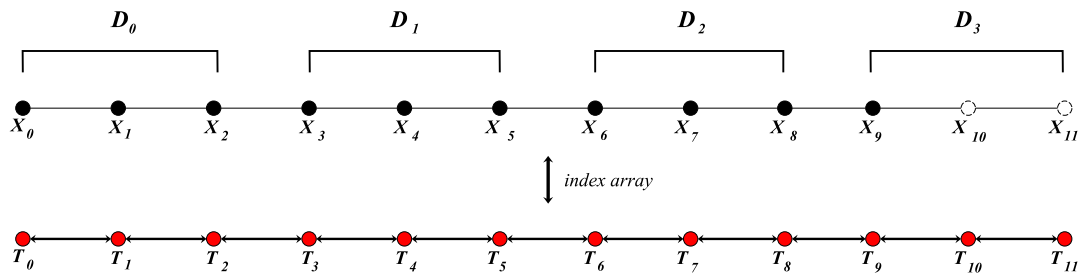}
    \centering
    \caption{The multi-GPU version of replica level parallelism combined with
    temperature swapping.}
    \label{fig_replica_level_parallelism}
\end{figure}

Replica exchanges that occur in the same GPU
are efficient since the swap can just be an exchange of pointers. In the limit
cases at $D_i^L$ and $D_i^R$, the task is not local to a single GPU anymore, 
since the exchange process would need to access replicas $D_{i-1}^R$ and
$D_{i+1}^L$, both which reside in different GPUs. Because of this special case, 
the pointer approach is not a robust implementation technique for a multi-GPU 
approach, neither explicit spin exchanges because it would require several memory
transfers from one GPU to another.  The solution to this problem is to swap
temperatures, which is totally equivalent for the result of the simulation. One just
needs to keep track of which replica has a given temperature and know their
neighbors. For this we use two index arrays, $trs$ and $rts$, for
\textit{temperatures replica sorted} and \textit{replica temperature sorted},
respectively. Figure \ref{fig_replica_level_parallelism} illustrates the whole
multi-GPU approach. 

The temperature swap approach is an efficient technique for single as well as
for clusters of GPU-based workstations, since for each exchange only two
floating point values (\textit{i.e.}, $2 \times 32 bits$) need to be swapped. 
This part of the algorithm results in little overhead compared to the simulation time, 
thus we opted to implement it on the CPU side.

\section{Adaptive Temperatures}
\label{sec_ATS}
The random field, as any other system with quenched disorder, becomes difficult
to study as $L$ increases. For $L \ge 64$ the selection of parameters is already
a complex task, since a small change on one can lead to a bad quality
simulation. 

One important parameter for simulation is the selection of temperatures and the
distance among them. In general, for the low-temperature regime and near the
transition point $T_c$, the replicas need to be placed at temperatures much
closer compared to the high temperature regime in order to achieve an acceptable
exchange rate. If these temperature requirements are not met, the simulation may
suffer from exchange bottlenecks, preventing the information to travel from one
side to other. Indeed one can decide to simulate with a dense number of
temperatures for the whole range, but this strategy tends to be inefficient
because the simulation does much more work than it should.  Based on this facts,
we propose to use an adaptive method that chooses a good distribution of
temperatures based on the exchange rates needed at each region.

The idea is to start with $M$ replicas, equally distributed from $T_{low}$
to $T_{high}$. The adaptive method performs an arbitrary number of trial
simulations to measure the exchange rate between each consecutive pair of
replicas. After each trial simulation, an array of exchange rates is obtained
and put in a min-heap. For the $a$ intervals with lowest exchange rate
(with $a$ chosen arbitrarily), new replicas are introduced using its middle
value as the new temperature.  
\begin{figure}[ht!]
    \includegraphics[scale=0.24]{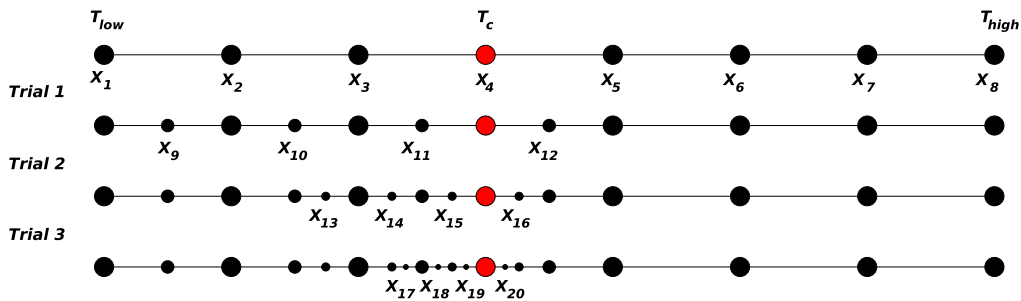}
    \centering
    \caption{The adaptive temperature strategy choosing the lowest four
    exchange rates for three trial runs.}
    \label{fig_adaptive}
\end{figure}

The trial simulations use multi-GPU computation.  Given that the
number of trial runs and the number of intervals to split is given a priori, the
memory pool can be allocated before hand and equally distributed among all
GPUs. After each trial run, the new $i$-th new replica is assigned to the $j$-th GPU with
$j = i \mod k$, where $k$ is the number of GPUs. The order in which new replicas
are assigned to each GPU actually does not affect the performance neither the
result of the simulation.

\section{Performance results}
\label{sec_multigpu_performance}
In this section we present the performance results of our multi-GPU implementation (which
we have named \textit{trueke} for \textit{exchange} in Spanish, which is
\textit{trueque}) and compare them against other
GPU and CPU implementations. The purpose of including a comparison against a
sequential CPU implementation is not to claim very high speedups, but rather to
provide a simple reference point for future comparisons by the community.

\subsection{Benchmark Plan}
Four performance metrics, averaged over $N$ repetitions, are used to measure the parallel performance of
\textit{trueke} (Note, in this section $T$ means running time, not temperature):
\begin{enumerate}
\item{Benchmark the \textbf{Spin-level Performance} by computing the average time of a spin flip:
    \begin{equation}
    \langle T_{spin} \rangle = \frac{1}{L^3 N \cdot w}\sum_{i=1}^{N}T_w
    \end{equation}
    where $w$ corresponds to the number of sweeps chosen for the Metropolis simulation.
}
\item{Benchmark the \textbf{Replica-level Performance} by computing the average time of a parallel tempering realization:
    \begin{equation}
    \langle T_{rep} \rangle = \frac{1}{N \cdot x}\sum_{i=1}^{N}T_{x}
    \end{equation}
    where $x$ is the number of exchange steps chosen for the simulation.
}
\item{Benchmark the \textbf{Multi-GPU Performance Scaling} by measuring, based on the
single GPU and multi-GPU running times $t_1, t_g$, the fixed-size speedup $S_{GPU}$ and
efficiency $E_{GPU}$
    \begin{equation} 
        \begin{split}
            S_{GPU} & = T_{1}/T_{g}\\
            E_{GPU} & = S_{GPU}/g
        \end{split}
    \end{equation}
    at different problem sizes when using two ($g=2$) GPUs.
}
\item{Benchmark the \textbf{Adaptive Temperatures Performance} by computing the average time 
of an adapted parallel tempering realization:
    \begin{equation}
    \langle T_{adapt} \rangle = \frac{1}{N}\Bigg(\sum_{i=1}^{N}T_k + \sum_{i=1}^{N}T_{x}\Bigg)
    \end{equation}
}
where $T_k$ is the time for doing $k$ trials.
\end{enumerate}
The number of repetitions (\textit{i.e.}, $N$) for computing the performance averages
range between $[10, 40]$ (\textit{i.e.}, less repetitions are required for
benchmarking large problem sizes), which are sufficient to provide a standard error
of less than $1\%$.

The workstation used for all benchmarks (including the comparison implementations)
is equipped with two 8-core Intel Xeon CPU E5-2640-V3 (Haswell), 128GB of RAM and two
Nvidia Tesla K40 each one with 12GB.

\subsection{Spin and Replica level Performance Results}
The first two benchmark results are compared, for reference, against a cache-aligned CPU
implementation of the 3D Ising Random Field running both sequential and multi-core. 
Additionally, we include the Spin-level performance of Weigel's GPU implementation
for the 2D Ising running on our system, as a reference of how close or far trueke's spin flip
performance is compared to Weigel's highly optimized code for 2D Ising model. The $L$ values
used in Weigel's implementation, which simulate $L^2$ spins, have been adapted to the
form $L' = \sqrt[3]{L^2}$ so that the input sizes can be compared against the 3D
ones, in the number of total spins.

\begin{figure*}[ht!]
\centering
\begin{tabular}{cc}
\includegraphics[scale=0.57]{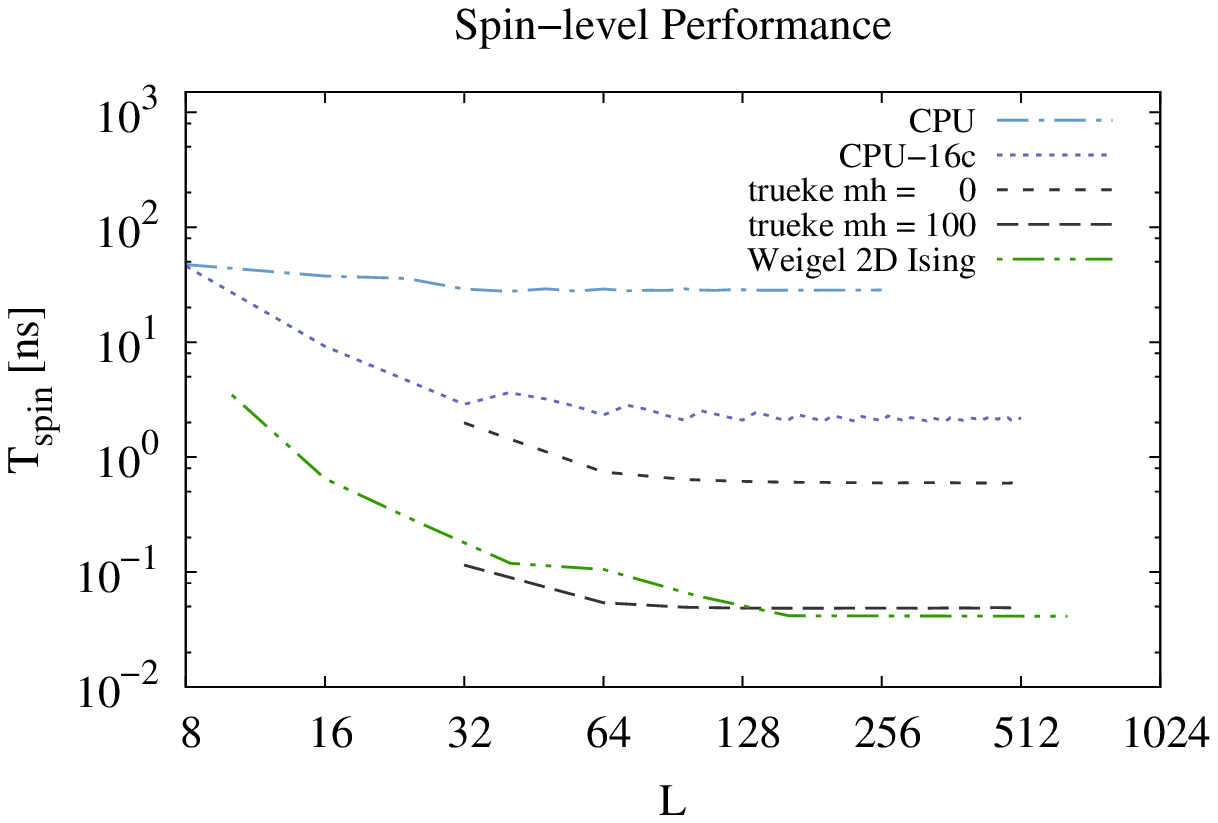} &
\includegraphics[scale=0.57]{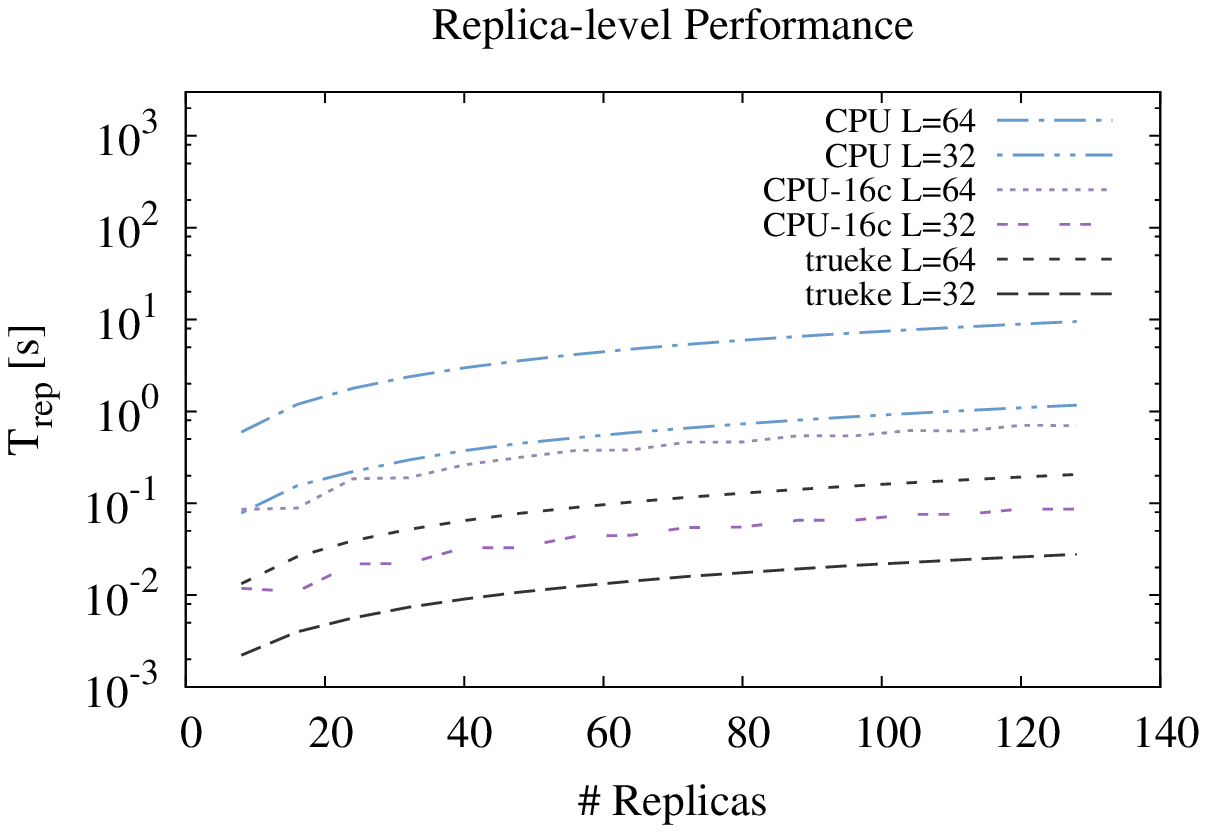}
\end{tabular}
\caption{On the left, spin-level performance. On the right, replica-level performance.}
\label{fig_levelperf}
\end{figure*}

Figure \ref{fig_levelperf} presents the results of spin flip time and the average
parallel tempering time. 

On the spin level performance results (left plot of Figure \ref{fig_levelperf}), we
observe that \textit{trueke} with $100$ multi-hit updates (\textit{i.e.}, $mh = 100$)
is over two orders of magnitude faster than the sequential CPU implementation and
over one order of magnitude faster than the multi-core CPU implementation using 16
cores.  It is worth noting that \textit{trueke} is almost as fast as Weigel's highly
optimized GPU Metropolis implementation for the 2D Ising model which uses $mh = 100$
too. If no multi-hit updates are used, then the performance of trueke decreases
as expected, becoming $6\times \sim 7\times$ faster than the multi-core
implementation. It is important to consider that the comparison has been done using
just one GPU for trueke while using two CPU sockets (8 cores each) for the CPU
implementation. Normalizing the results to one silicon chip, one would obtain that
for spin-level performance the GPU performs approximately one order of magnitude
faster than a multi-core CPU. In order to obtain good quality physical results, the
3D Ising Random Field model must be simulated with $mh = 0$.  For this reason, the
rest of the results do not include the case when $mh = 100$.

The replica performance results (right plot of Figure \ref{fig_levelperf}) shows that
the multi-GPU implementation outperforms the CPU implementation practically in the
same orders of magnitude as in the spin level performance result with $mh = 0$. This
result puts in manifest the fact that the exchange phase has little impact on the replica
level parallelism, indicating that multi-GPU performance should scale efficiently.

\subsection{Multi-GPU Scaling}
\begin{figure*}[ht!]
\centering
\begin{tabular}{cc}
\includegraphics[scale=0.57]{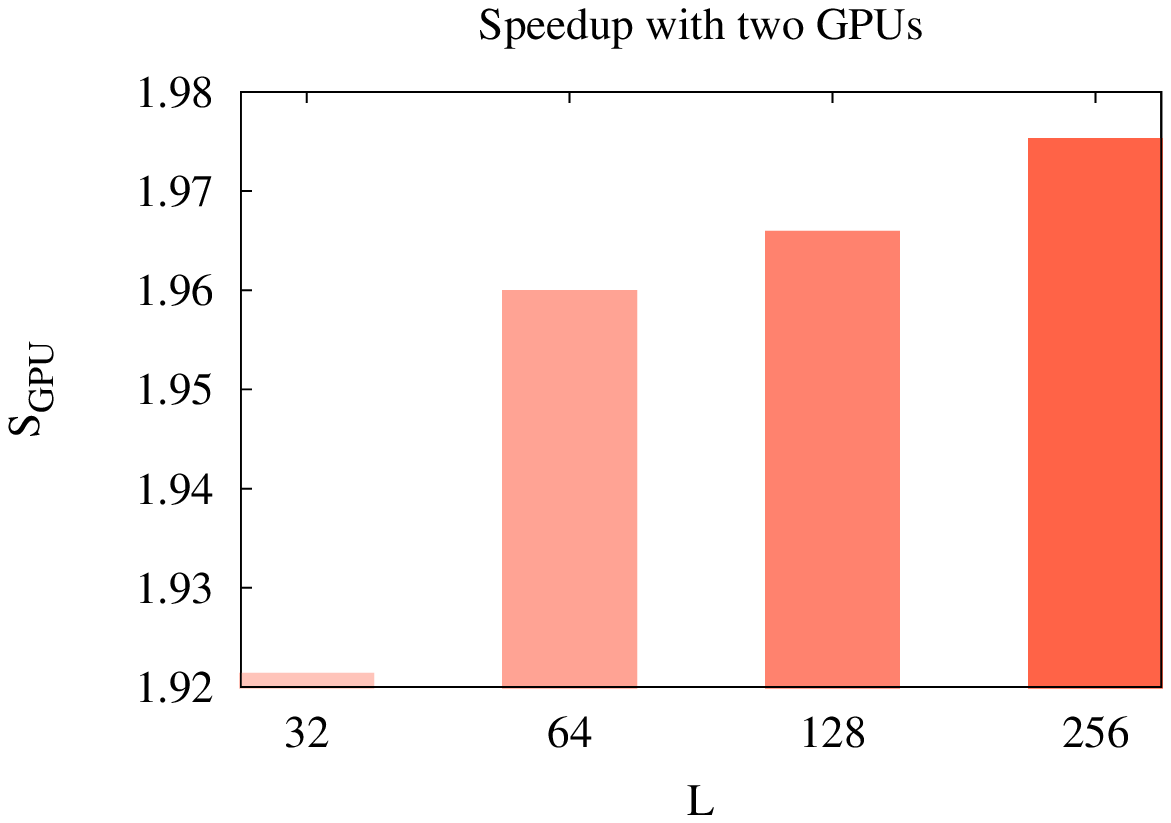} &
\includegraphics[scale=0.57]{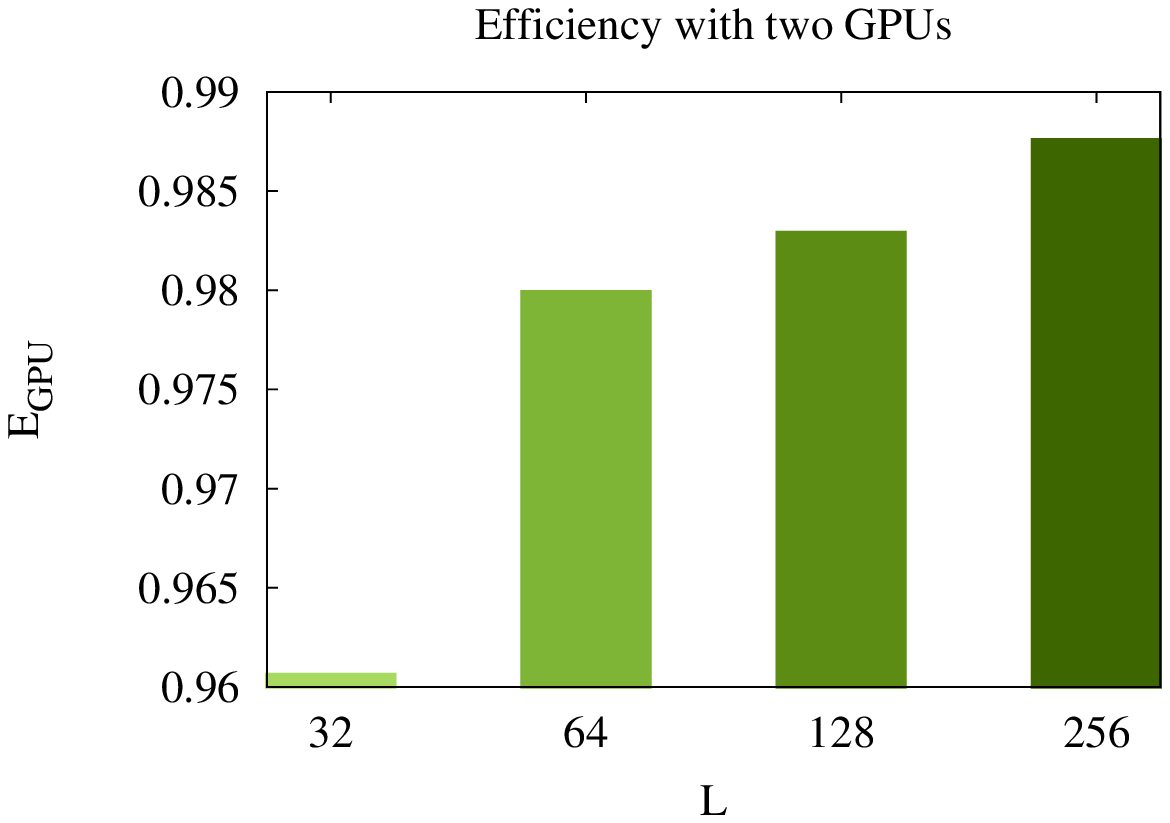}
\end{tabular}
\caption{Multi-GPU speedup and efficiency for different lattice sizes.}
\label{fig_scaling}
\end{figure*}

Figure \ref{fig_scaling} presents the speedup and efficiency of 
\textit{trueke} for computing one 3D Ising Random Field simulation 
using two Tesla K40 GPUs.

From the results, we observe that as the problem gets larger, the replica level
speedup improves almost to $S_{GPUs=2} = 2$ which is the perfect linear speedup. The
efficiency plot shows that the replica-level parallel efficiency approaches to
$E_{GPUs=w} = 1$ as the lattice becomes larger, indicating that replica-level
parallelism does not degrade when increasing $L$. This favorable behavior can be
explained in part by the fact that the work of the exchange phase grows at least as
$W_{ex} = \Omega(M) + \Omega(ML^3)$ where the $\Omega(M)$ term is the sequential work
for exchanging the $M$ replicas, while the $\Omega(ML^3)$ term is the parallel work
for computing the energy at each replica, which is done with a parallel GPU
reduction, in $O(log(L^3))$ time for each replica. It is clear that the amount of
parallel work grows faster than the sequential work, therefore the parallel
efficiency of the whole method should be higher as $L, M$ and the number of GPUs
increase. 

\subsection{Performance of The Adaptive Temperatures Technique}
Table \ref{table_adaptperf} presents the running times of the \textit{adaptive
temperatures} technique compared to dense and spare homogeneous techniques for $L =
32, 64$. The simulation parameters used for all simulations were $100$ disorder
realizations, each one with $2000$ parallel tempering steps and $10$ Metropolis
sweeps.  For $L=32$, the adaptation phase used $10$ trial runs with $2$ insertions at
each trial. For $L=64$, the adaptation used $32$ trial runs, with $3$ insertions at
each trial. The trial runs also use $2000$ parallel tempering steps with $10$
Metropolis sweeps.

\begin{table}[h!b!p!]
\begin{center}
\caption{Executions times, in seconds, for adaptive and homogeneous approaches.}
  \begin{tabular}{ c | c | c | c | c | c}
    \hline
	L	& sparse-sim    & dense-sim   &  adapt-trials  & adapt-sim & adapt (trials + sim)  \\ 
	\hline
	32 	&	670.25 		&	1417.04   &  41.84         & 1167      & 1209.49  \\
    64 	&	10386.75	&   26697.17  &  1689.46       & 19994.70  & 21684.15 \\
  \end{tabular}
  \label{table_adaptperf}
\end{center}
\end{table} 

From the results, it is observed that a full adaptive simulation, including its
adaptation time, is more convenient than a dense simulation with no adaptation from
the point of view of performance. The sparse technique is the fastest one because it
simply uses less replicas, but as it will be shown in the next Section, it is not a
useful approach for a disordered system starting from $L \ge 64$, neither the dense
one, because the exchange rate becomes compromised at the low temperature regime.

\section{Exchange Rates with Adaptive Temperatures}
\label{sec_multigpu_adaptive_exchange}
\begin{figure*}[ht!]
\centering
\begin{tabular}{cc}
\includegraphics[scale=0.57]{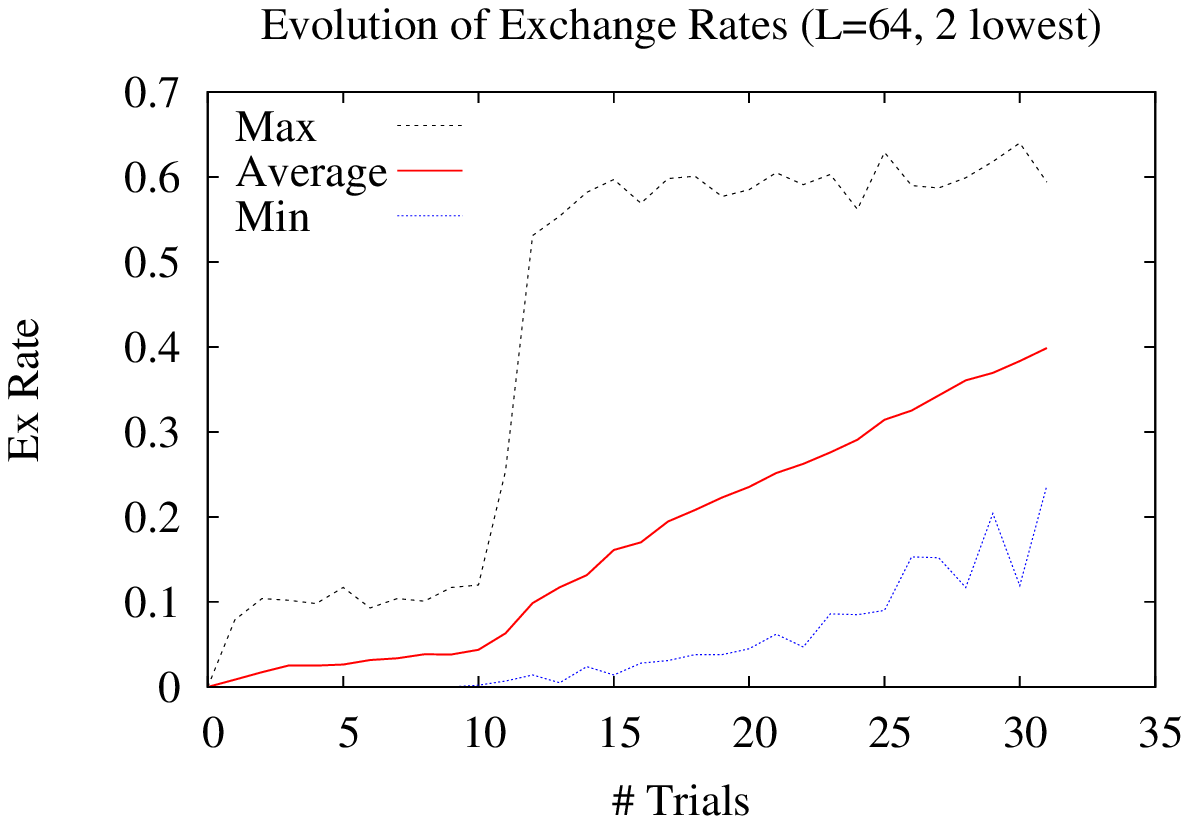} & 
\includegraphics[scale=0.57]{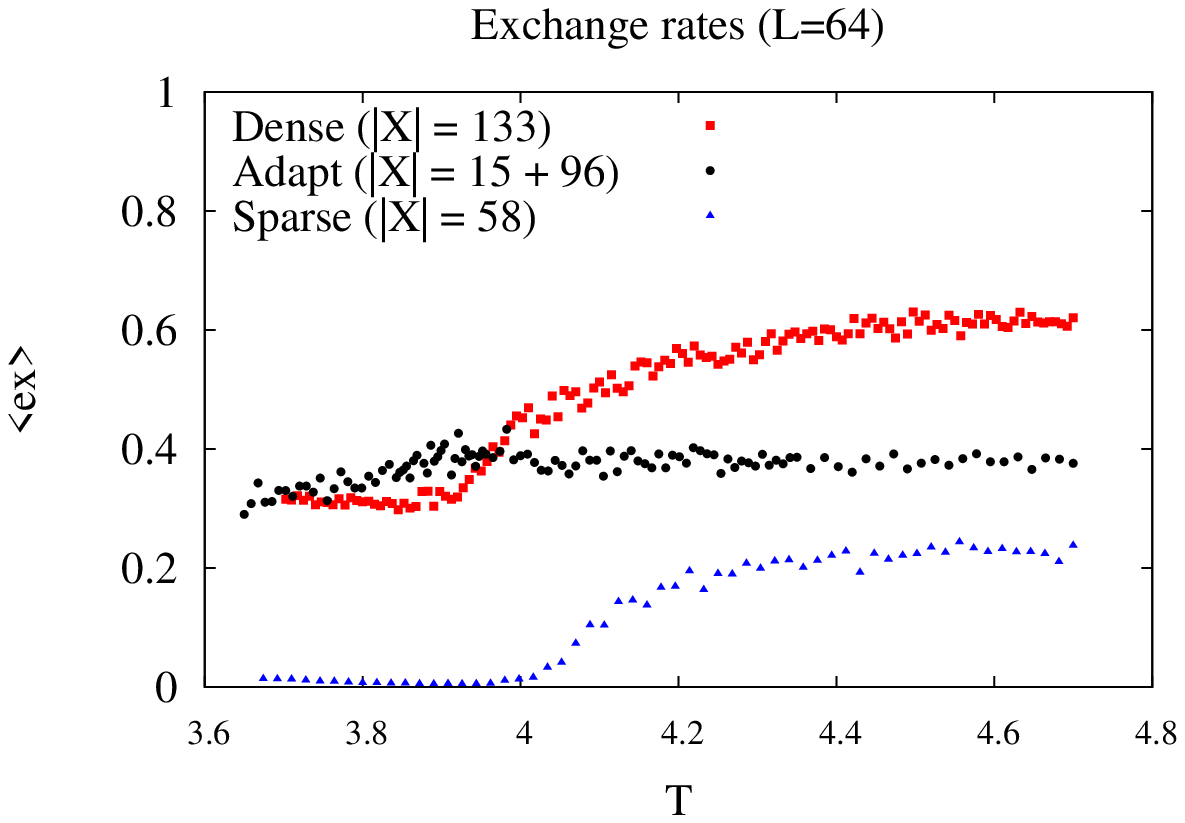}
\end{tabular}
\caption{On the left, the evolution of average, min and max exchanges through the
trial runs. On the right, the exchange rates for the whole temperature range.}
\label{fig_adapt}
\end{figure*}

Results on the exchange rate of the adaptive strategy is presented by plotting the
evolution of the average, minimum and maximum exchange rates through the trial runs.
Also a comparison of the exchange rates between a dense homogeneous set, the
adaptive set and a sparse homogeneous set is presented.  The simulation parameters
were the same as the ones used to get the performance results of Table
\ref{table_adaptperf}. Figure \ref{fig_adapt} presents the results. 

From the left plot we observe that as more replicas are added, the average and
minimum exchange rates increase and tend to get closer, while the maximum exchange
rate stabilizes after starting from trial number $10$. For the right plot, it is
clearly shown that the dense approach, although is an homogeneous approach, does not
generate an homogeneous exchange rate value for the whole temperature range. For the
sparse approach the scenario is even worse because at the low temperature regime
there almost no exchanges. On the other hand, the adaptive method generates an
exchange rate that is almost homogeneous for the entire range, which is preferred in
order to have more control of the simulation. The numbers on the labels indicate the
number of replicas used, and it can be seen that the adaptive method uses less
replicas than the dense homogeneous approach, therefore it runs faster.

\section{Preliminary Physical Results}
\label{sec_physical}
For the correctness of the whole algorithm, including the exchange phase, we ran
simulations in the 3D Random Field Ising Model with field strength $h=1$.  The
observables have been measured using $5000$ parallel tempering steps, $10$ sweeps at
each step, $1$ measurement at each parallel tempering step and using the adaptive
temperatures technique. 

\begin{figure*}[ht!]
\centering
\begin{tabular}{cc}
\includegraphics[scale=0.57]{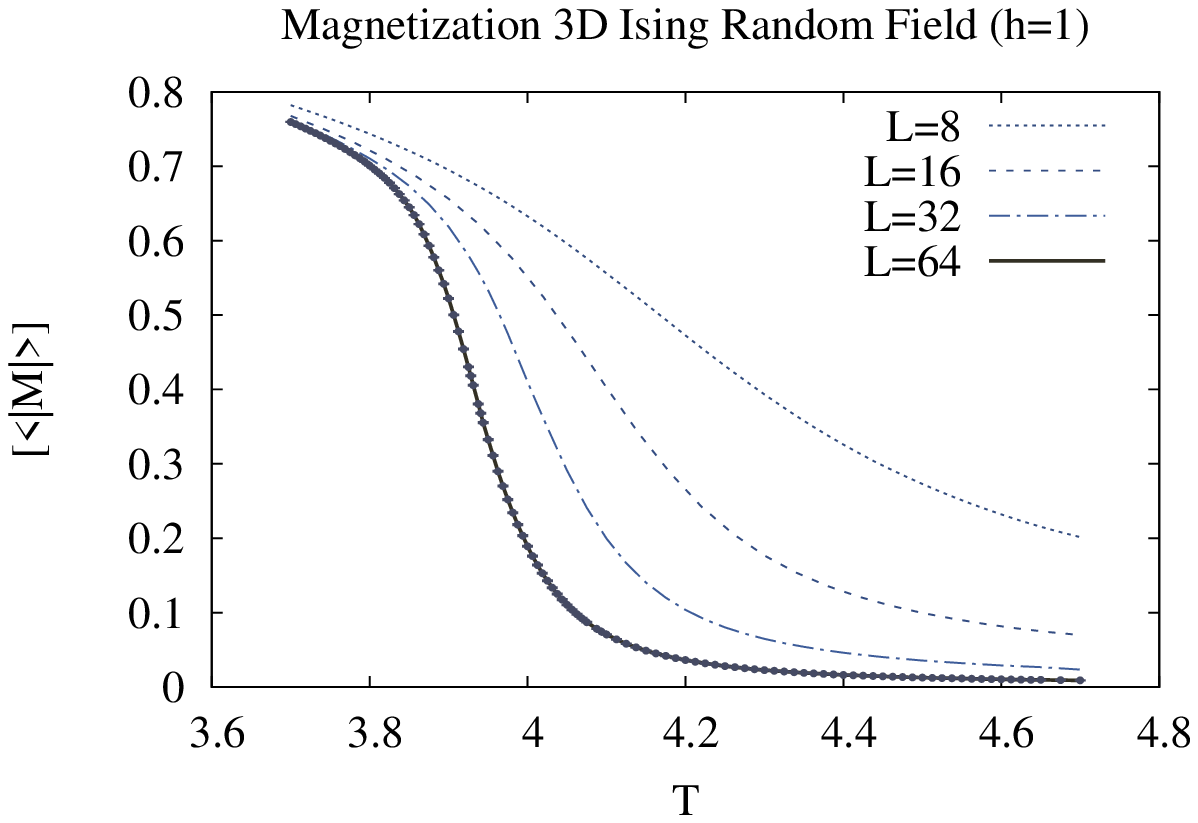} &
\includegraphics[scale=0.57]{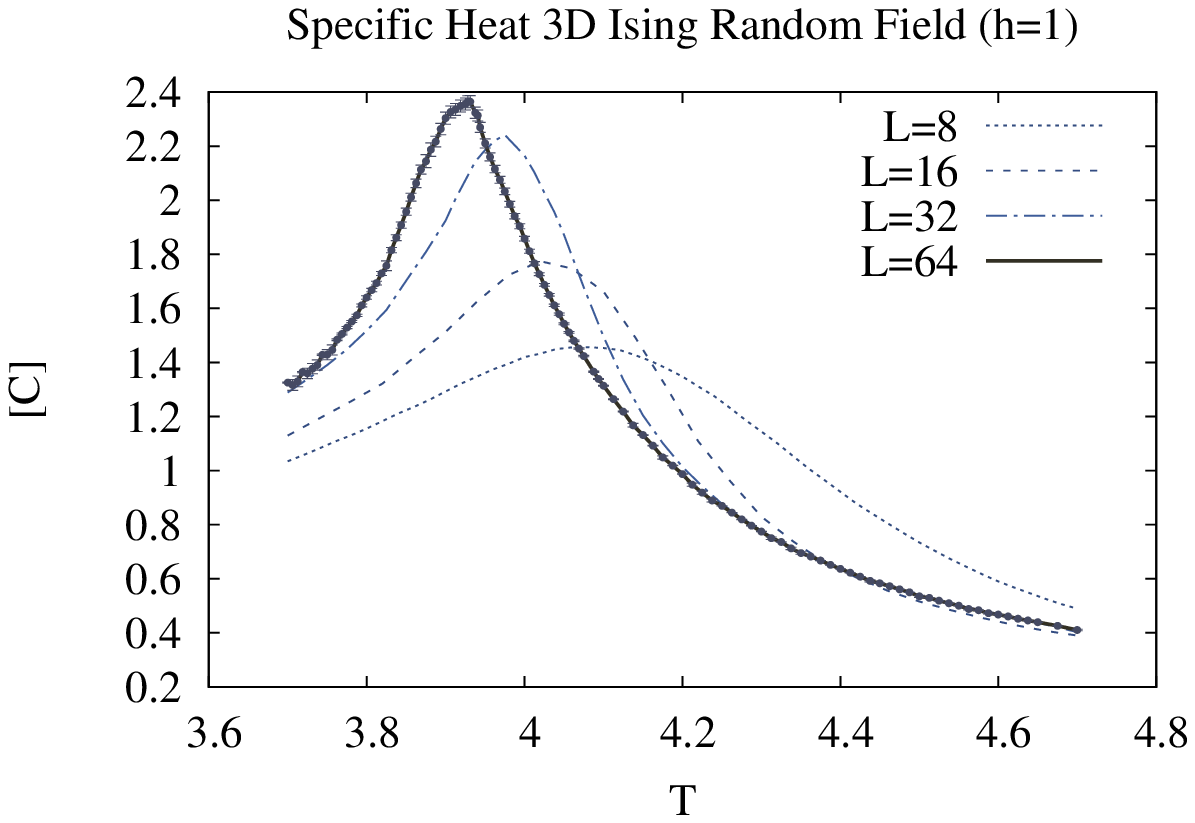}\\
\includegraphics[scale=0.57]{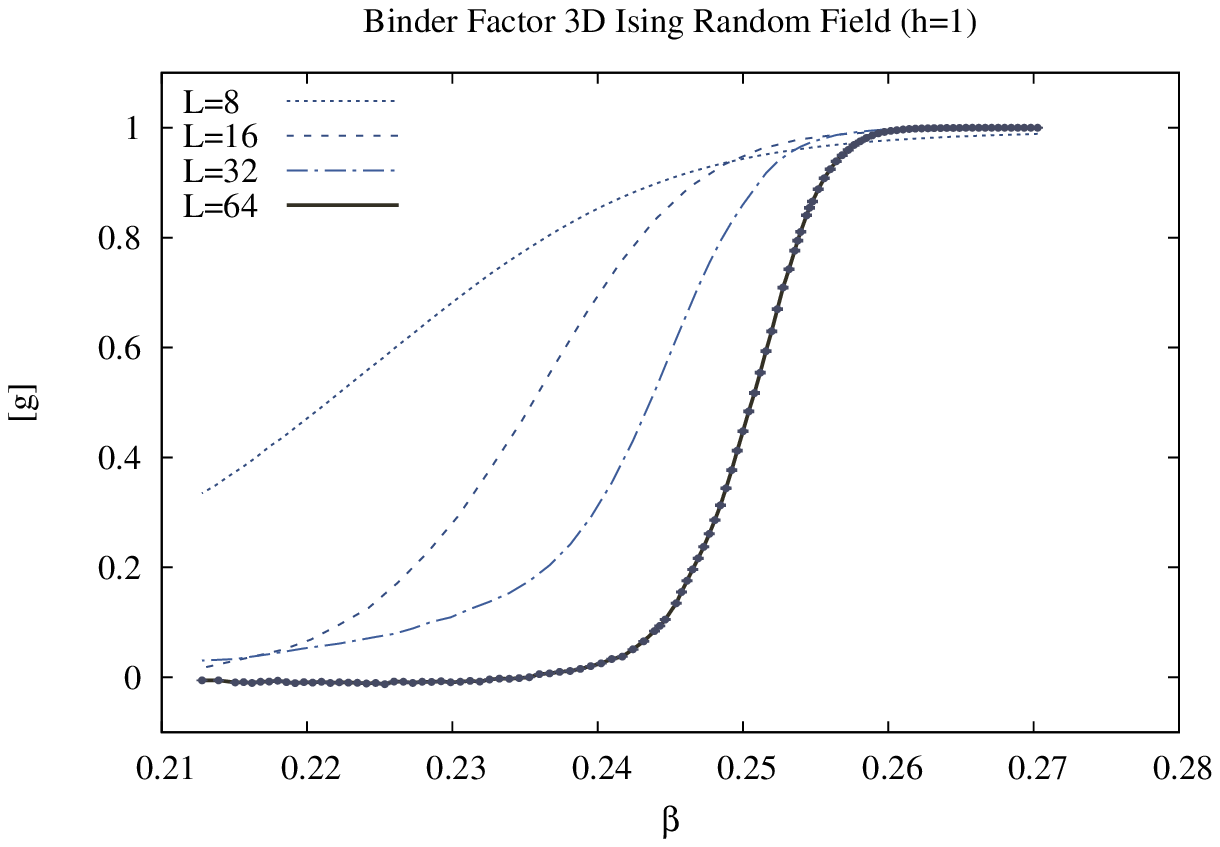} &
\includegraphics[scale=0.57]{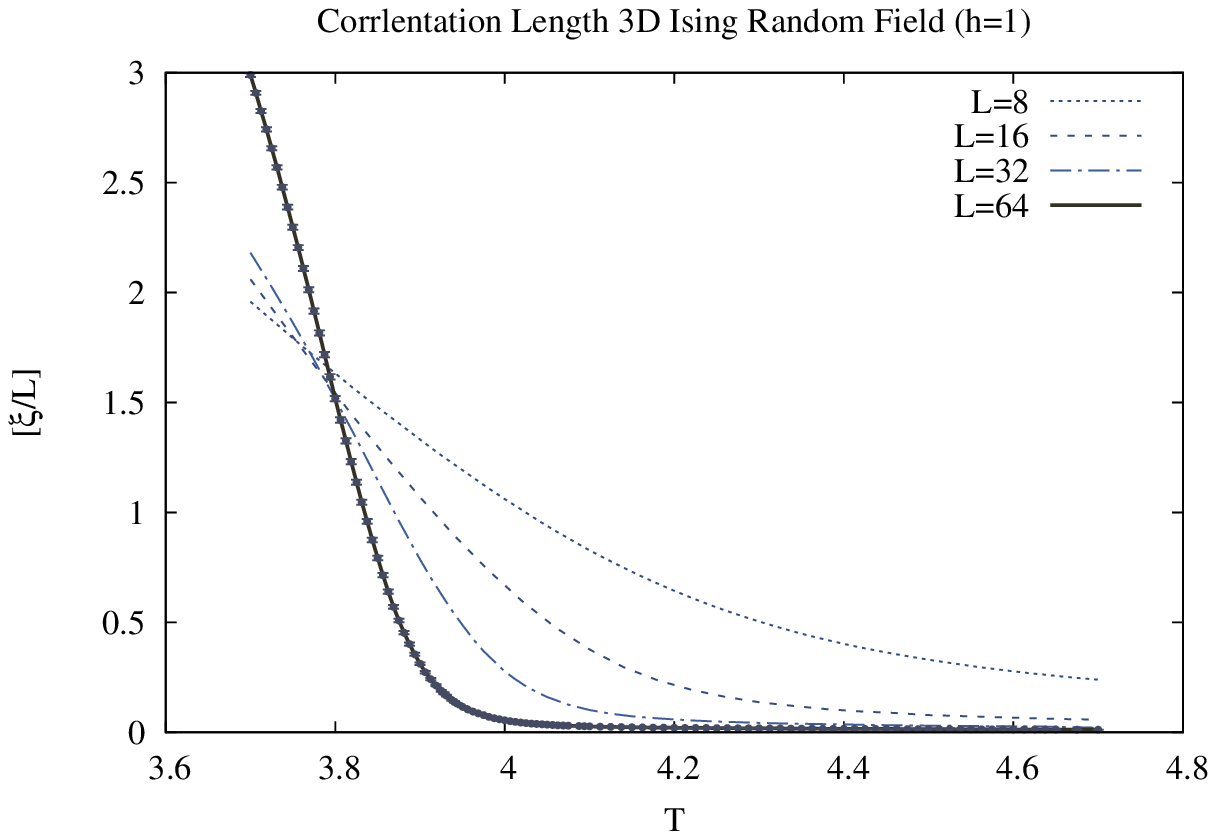}
\end{tabular}
\caption{Preliminary physical observables for the 3D Random Field with $h=1$.}
\label{fig_rfresults}
\end{figure*}

Average observables are computed as $[\langle A \rangle]$, where $[..]$ corresponds to
the average over different disorder realizations and $\langle A \rangle$ denotes the 
thermal average for a single random field configuration. 
The magnetization $\langle |M| \rangle$ is defined as
\begin{equation}
    [\langle |M| \rangle] = \Bigg [\Big \langle \Big| \frac{1}{V} \sum_{i=1}^{L^3} s_i \Big|
    \Big\rangle \Bigg]
\end{equation}
The specific heat is
\begin{equation}
    [C] = \frac{L^3}{T^2} \Big[ (\langle E^2 \rangle - \langle E \rangle^2) \Big]
\end{equation}
where $E$ is the average energy per site. The Binder factor is an average at the
disorder realization level and it is defined as:
\begin{equation}
    [g] = \frac{1}{2} \Big(3 - \frac{[\langle M^4 \rangle]}{[\langle M^2 \rangle]^2}\Big)
\end{equation}
and the correlation length is
\begin{equation}
    [\xi] = \Big[\sqrt{\frac{\langle M^2 \rangle}{\langle F \rangle}- 1}\Big]
\end{equation}
with:
\begin{alignat}{2}
    F    &= \frac{1}{3L^3} \Big(F_1 + F_2\Big)\\
    F_1  &= \Big(\sum_{i}^{L^3} h_{i} \cos(K\cdot i_x)\Big)^2 + \Big(\sum_{i}^{L^3} h_{i} \cos(K\cdot i_y)\Big)^2 + \Big(\sum_{i}^{L^3} h_{i} \cos(K\cdot i_z)\Big)^2\\
    F_2  &= \Big(\sum_{i}^{L^3} h_{i} \sin(K\cdot i_x)\Big)^2 + \Big(\sum_{i}^{L^3} h_{i} \sin(K\cdot i_y)\Big)^2 + \Big(\sum_{i}^{L^3} h_{i} \sin(K\cdot i_z)\Big)^2
\end{alignat}
where $K = 2\pi / L$,  $h_i = \{-1,1\}$ is the disordered magnetic field value at
spin $s_i$ and $\{i_x, i_y, i_z\}$ correspond to the spatial coordinates of a given
spin $s_i$ in the lattice.  For visual clarity, we only included the error bars of
the largest size studied, \textit{i.e.}, $L=64$, nevertheless it is worth mentioning
that the error bars for $L= 8, 16, 32$ were even smaller than the ones for $L = 64$.
The results are presented in Figure \ref{fig_rfresults} and confirm the
transition-like behavior at $3.8 \le T_c \le 3.9$, or $0.2564 \le \beta_c \le
0.2631$, as shown by Fytas and Malakis in
their phase diagram when $h = 1$ \cite{RFIM2008}. 
The results presented in this section are preliminary, with up to $2000$ disorder
realizations for $L=64$ (less for $L < 64$), each doing $5000$ exchange Monte Carlo steps 
with $10$ Metropolis-Hastings sweeps between exchanges.  A physical
paper devoted to the physical results will be prepared for the future. 

\section{Discussion}
\label{sec_multigpu_conclusions}
We presented a multi-GPU implementation of the \textit{Exchange Monte Carlo} method,
named \textit{trueke}, for the simulation of the 3D Ising Random Field model. The
parallelization strategy is organized in two levels; (1) \textit{spin-level
parallelism}, which scales in the presence of more cores per GPU, and (2) the
\textit{replica-level parallelism} that scales in the presence of additional GPUs.
The spin-level parallelism is up to two orders of magnitude faster than its CPU
counterpart when using multi-hit updates, and between one and two orders of magnitude
faster when not using multi-hit updates. The parallel scaling of the method improved
as the problem size got larger, in part because the amount of parallel work increases
faster than the sequential work (\textit{i.e.}, exchange phase). This behavior is
indeed favorable for multi-GPU computation, where \textit{trueke} achieved
approximately up to $99\%$ of efficiency when using two GPUs.  Due to hardware
limitations, we could not go beyond two GPUs, which would have been ideal for having
a better picture of how the performance would scale in large systems. Nevertheless,
we intend to put \textit{trueke} available to the community in the near future, so
that it can be benchmarked in systems with a high number of GPUs. 

The adaptive strategy for selecting the temperature set was based on the idea of
inserting new temperatures in between the lowest exchange rates found by an arbitrary
number of trial runs. As a result, the simulation used more replicas at the locations
were exchange rates were originally low, and less replicas were the exchange rate was
already good, such as in the high temperature regime. The adaptive strategy
outperformed any homogeneous approach, since these last ones had to deal with an
over-population of replicas at places that actually did not require more temperature
points, resulting in extra computational cost and slower performance, and an
under-population of replicas at the low temperature regime near $T_c$, generating low
exchange rates. The adaptive method works better when a
small number of points are added at each trial run, \textit{i.e.}, between one and
five insertions at each trial run, because this way is more unlikely to misplace an
insertion. Compared to the method by Katzgraber \textit{et. al.}
\cite{1742-5468-2006-03-P03018}, our adaptive method differs since it feedbacks from the local
exchanges of pairs of temperatures, always lifting the minimum values observed by
inserting new temperatures, while in the work of Katzgraber \textit{et. al.} they
feedback by counting the number of times a replica travels the whole temperature
range and based on this information they move the temperature set. In the method by
Bittner \textit{et.  al} \cite{PhysRevLett.101.130603}, they vary the number of
Metropolis sweeps based on the auto-correlation times in order to avoid two replicas
getting trapped exchanging together, which they identify as a problem for Katzgraber
method. Our approach of inserting new temperatures provides the advantage that it
does not compromise the rest of the temperature range and the technique by itself is
general since it is just based on improving the lowest $k$ minimum exchange rates
observed at each trial run. For the near future, we intend to do a formal
comparison of all three implementations of the methods.

The main reason for choosing multi-GPU computing at the replica-level, and not at the
realization level (fully independent parallelism) as one would naturally choose, is
mostly because the latter strategy is not prepared for the study of large lattice
systems, which would be of great interest for the near future. From our experience
with the 3D RFIM, the number of replicas needed to keep at least $35\% \sim 45\%$ of
exchange rate grew very fast as $L$ increased. Thus, distributing the replicas
dynamically across several GPUs extends the possibilities of studying larger
disordered lattices. 

The multi-GPU approach proposed, alias \textit{trueke}, has allowed us to study th 3D
Ising Random Field model at size $L = 64$ for which its results can become physical
contributions to the field. It is expected that eventually, as more GPUs are used, 
larger lattices could be studied. 

\section{Acknowledgements}
The authors would like to thank Prof. M. Weigel for his valuable feedback and
explanations on the subject. This research has been supported by the PhD program from
CONYCIT, Chile.  This work is also supported by the National Science Foundation of
China (NSFC) under Grant No. 11275185 and by the Open Project Program of State Key
Laboratory of Theoretical Physics, Institute of Theoretical Physics, Chinese Academy
of Sciences, China (No. Y5KF191CJ1). 
\bibliographystyle{abbrv}
\bibliography{\bibtexdb}

\end{document}